\documentclass[aps,prl,twocolumn,superscriptaddress]{revtex4-1}

\usepackage{graphicx}
\usepackage{epstopdf}
\usepackage[T1]{fontenc}

\newcommand{\bvec}[1]{\mbox{\boldmath $#1$}}

\begin{document}

\title{Valence-selective local atomic structures on a YbInCu$_4$ valence transition material by x-ray fluorescence holography}

\author{Shinya~Hosokawa}
\email[Corresponding Author: ]{shhosokawa@kumamoto-u.ac.jp}
%\thanks{}
\affiliation{Department of Physics, Kumamoto University, Kumamoto 860-8555, Japan}

\author{Naohisa~Happo}
\affiliation{Graduate School of Information Sciences Hiroshima City University, Hiroshima 731-3194, Japan}

\author{Kouichi~Hayashi}
\author{Koji~Kimura}
\affiliation{Department of Physical Science and Engineering Nagoya Institute of Technology, Nagoya 466-8555, Japan}

\author{Tomohiro~Matsushita}
\affiliation{Japan Synchrotron Radiation Research Institute (JASRI), Sayo 679-5198, Japan}

\author{Jens~R\"{u}diger~Stellhorn}
\thanks{Present address: Deutsches Elektronen-Synchrotron (DESY), 22603 Hamburg, Germany}
\affiliation{Department of Physics, Kumamoto University, Kumamoto 860-8555, Japan}

\author{Masaichiro Mizumaki}
\author{Motohiro Suzuki}
\affiliation{Japan Synchrotron Radiation Research Institute (JASRI), Sayo 679-5198, Japan}

\author{Hitoshi~Sato}
\affiliation{Hiroshima Synchrotron Radiation Center, Hiroshima University, Higashi-Hiroshima 739-0046, Japan}

\author{Koichi~Hiraoka}
\affiliation{Graduate School of Science and Engineering, Ehime University, Matsuyama 790-8577, Japan}

\date{\today}

\begin{abstract}
An experimental technique of x-ray fluorescence holography for investigating valence-selective local structures was established by employing the incident x-ray energy at a characteristic one near an x-ray absorption edge, and it was adopted to discriminate environments around Yb$^{2+}$ and Yb$^{3+}$ ions in a YbInCu$_4$ valence transition material below and above the transition temperature of 42 K. The reconstructed images of the neighboring atoms around Yb$^{3+}$ show a $fcc$ structure as observed by diffraction experiments, whereas those around Yb$^{2+}$ have curious cross (+) features, indicating a positional shift of the centered Yb$^{2+}$ ions from the $fcc$ lattice point.
\end{abstract}

\maketitle

%\section{Introduction}
An abrupt change was reported by Felner and Nowik \cite{Felner86} in 1986 in the temperature ($T$) dependence of magnetic susceptibility of Yb$_x$In$_{1-x}$Cu$_2$ ($x\sim0.3-0.6$). A simple valence-fluctuation model was proposed, by which a first-order Yb$^{3+}$ to Yb$^{2+}$ phase transition was predicted with simply decreasing $T$. These compounds exhibit the sharpest $T$-dependent valence phase transition in any metallic systems.

Subsequent to this finding, Felner et al. \cite{Felner87} exhibited several physical properties on this transition in mainly Yb$_{0.4}$In$_{0.6}$Cu$_2$ alloy. An x-ray diffraction (XRD) experiment reveals a cubic Laves structure with the space group of $Fd\bar{3}m$ at all the temperatures from 300 to 4.2 K with a sudden increase in the unit cell size by about 0.15\% at 60-40 K ($T_v$ at $x=0.4$). A neutron diffraction measurement proves the absence of magnetic order. An $^{170}$Yb M\"{o}ssbauer study shows that the Yb ion is magnetic beyond $T_v$ at 60 K, whereas it is non-magnetic below $T_v$ at 4.2 K. An electrical resistivity exhibits a large decrease by about 25\% from 60 to 40 K. A specific-heat measurement reveals a characteristic increase around $T_v$. Finally, An x-ray absorption near edge structure (XANES) study at the Yb $L_{\rm III}$ edge shows a change in the $4f$-electron occupancy at $T_v$, and it was found that the valences of 2.9 and 2.8 above and below $T_v$, respectively, are independent of $T$ in these temperature regions. In addition, substitution, pressure, and magnetic-field dependences of the valence phase transition were investigated by the same group in detail \cite{FelnerJMMM,Nowik}.

Then, the interests have moved to YbInCu$_4$ ($x=0.5$) because the sample of $x=0.4$ is primarily in two phases of YbInCu$_4$ and InCu$_2$, which was found by Kojima et al. \cite{KojimaJMMM} using an electron-probe micro-analysis observation. They measured XRD and found the cubic C15b crystal type with the space group of $F\bar{4}3m$, in which Yb and In ions are ordered in a $fcc$ site. The resistivity of a well-annealed sample shows a sudden change around 40-45 K, which is narrower than the previous work for Yb$_{0.4}$In$_{0.6}$Cu$_2$ \cite{Felner87}. Their magnetic susceptibility and $^{115}$In Knight shift results showed a sharp valence phase transition occurs at $T_v=40$ K and Yb ions are in the valence fluctuating state below $T_v$. The structure was refined by powder neutron diffraction (ND) data \cite{KojimaJPSJ}, and the lattice constant shows an abrupt change of 0.15\% as was the previous study \cite{Felner87} at $T_v=40$ K without any change in the crystal symmetry. 

Physical properties on a single crystal YbInCu$_4$ were measured by Kindler et al. \cite{Kindler}. Mostly same values as the polycrystalline ones were obtained in the above parameters although $T_v$ is a higher value of 66.9 K. Owing to the single crystal sample, elastic constants can be measured in different directions, and a pronounced softening of the bulk modulus and an anomalous decrease in the Poisson ratio occur in the vicinity of $T_v$. Detailed comparisons of the structures of single and polycrystalline samples were carried out using neutron diffraction \cite{Lawrence}, and the sharpest single transition was found in the single crystal near 40 K. They also concluded that the neutron diffraction patterns measured above and below $T_v$ are identical to be C15b crystal type, and the increases of $T_v$ and gradual transitions are originated from a site-disorder between the Yb and In atoms. A similar argument was performed on a site-disorder between the In and Cu atoms by XRD using the $k$ dependence of x-ray atomic form factors \cite{Moriyoshi}. 

Concerning the $4f$-electron occupancy across $T_v$, the XANES analysis gave a small change of $2.98\rightarrow2.85$ \cite{Zhuang}. However, a bulk-sensitive hard x-ray electronic study of Yb $3d$ core-level and valence band photoemission measurements reveals a quite large change of $2.90\rightarrow2.74$ \cite{Sato}. The valency change was also observed by Yamaoka et al. \cite{Yamaoka} using bulk-sensitive methods of high-resolution x-ray absorption spectroscopy with partial fluorescence yields mode and resonant x-ray scattering spectroscopy at the Yb $L_{\rm III}$ absorption edge, and a change of approximately $2.94\rightarrow2.80$ was reported. Thus, a valence change of $\sim0.15$ is the consensus value for the valence transition, which causes large changes in physical properties as mentioned above. The charge transfer at $T_v$ was suggested to be between the Cu-derived conduction band and the Yb $4f$ states by Cu $2p_{3/2}$ soft x-ray absorption spectroscopy by Utsumi et al. \cite{UtsumiPRB}.

The atomic radius of Yb$^{2+}$ is larger than that of Yb$^{3+}$ by about 17\% even depending on the coordination numbers around the Yb ions \cite{Shannon}. So, the increase of the averaged atomic radii of Yb ions can be estimate to be $\sim2.5$\% on the valence transition, which is extremely larger than the increase of the lattice constants of $\sim0.15$\% at $T_v$. Thus, the Yb$^{2+}$ ions with the large atomic size should highly squeeze into a rigid crystal lattice below $T_v$, and large lattice distortions are expected around the Yb$^{2+}$ ions. 

A detailed XRD measurement using synchrotron radiation (SR) revealed an increase in the inter-Cu$_4$-tetrahedron below $T_v$ by about 0.2\% together with small increases of Yb-Cu and In-Cu distances, while intra Cu-Cu distance in the tetrahedron remains mostly unchanged \cite{UtaumiJJAP}. It would be an indirect structural effect for the change of Yb valency. Very recently, Tsutsui et al. found a small splitting of Bragg peaks in high-order reflections by a high-energy XRD experiment below $T-v$, and concluded that a structural change occurs from a cubic to a tetragonal phase at $T_v$ \cite{Tsutsui}, which contradicts the previous conclusions of unchanged crystal type \cite{Felner87,KojimaJMMM,KojimaJPSJ,Kindler,Lawrence,Moriyoshi,UtaumiJJAP}.

For the further structural investigation on the valence transition material, valence-selective method is essential. However, XRD has a small difference with valence, i.e., atomic form factors are slightly different owing to the electron numbers, and ND has no difference. X-ray absorption fine structure (XAFS) data shows an overlap of oscillations with different valencies with energy shifts. Therefore, a reliable valence-selective methods is highly required for evaluating the structural information on different valencies in YbInCu$_4$.

For this, we propose valence-selective x-ray fluorescence holography (XFH), which recently applied to Y oxide thin film \cite{Stellhorn} in which the valence is different between the surface and bulk, and to an Fe$_3$O$_4$ mixed valence material \cite{Ang}. XFH is a method for atom-resolved structural characterizations, and enables to draw three-dimensional (3D) atomic images around a specific element emitting fluorescent x-rays \cite{Tegze,HayashiJPCM}. When the incident x-rays have an energy higher than an absorption edge of an element, the target atom emits fluorescent x-rays. In addition, x-rays scattered by surrounding atoms also reach the target atom. The direct incident waves (reference wave) and the scattered waves (object wave) interfere each other, and the intensity of fluorescent x-rays is proportional to the interfered x-ray intensity. Thus, the fluorescent x-rays give a modulation of some 0.1\% with the incident x-ray angles with respect to the crystal axes of the sample, which is referred as a hologram. Then, the 3D images of the neighboring atoms can, in principle, be reconstructed $via$ simple Fourier transform-like approach with no special atomic models. Thus, XFH can observe local atomic arrangements in the short and intermediate ranges around the target atoms emitting fluorescent x-rays. 

\begin{figure}
\begin{center}
\includegraphics[width=75mm]{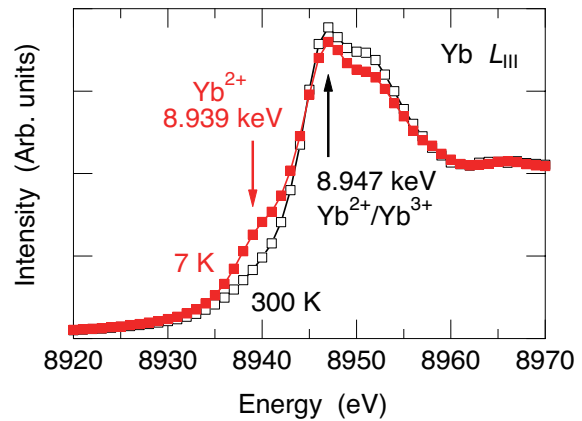}
\caption{\label{YbXANES}Yb $L_{\rm III}$ XANES spectra of YbInCu$_4$ at 7 K (closed squares) and 300 K (empty squares).}
\end{center}
\end{figure}

The important attempt for XFH to comprise the valence-selective character is to employ the incident x-ray energy, $E_0$, at an energy specific to a valence near the XANES region. Figure \ref{YbXANES} shows Yb $L_{\rm III}$ XANES spectra of YbInCu$_4$ used for the present study at 7 and 300 K measured in fluorescence mode, the feature of which is in very good agreement with the previous results \cite{Felner87} on Yb$_{0.4}$In$_{0.6}$Cu$_2$. At 7K, a shoulder is observed at about 8.939 keV, which is characteristic to the Yb$^{2+}$. When $E_0$ are employed at this energy, Yb $2p_{3/2}$ electrons in only Yb$^{2+}$ ions excite and emit the fluorescent x-rays, and those in Yb$^{3+}$ ions do not. Thus, the obtained hologram at this energy composes valence-selective structural information around mainly Yb$^{2+}$ ions.

%\section{Experimental procedure}
A single crystal YbInCu$_4$ was grown by a flux growth method. Constituent elements with stoichiometric ratios in InCu flux were put in an alumina crucible and sealed in an evacuated quartz ampoule. The sample was then heated to 1100$^\circ$C and cooled slowly down to 800$^\circ$C. After keeping at 800$^\circ$C for 20 h, the flux was removed. The crystal was cleaved so as to have a flat (001) surface with an area larger than $1\times1$ mm$^2$. The crystallinity of the sample was examined by taking a Laue photograph, and the concentration and homogeneity over the sample were confirmed within the experimental errors by an electron-probe microanalysis.

Yb $L\alpha$ XFH measurements were carried out at 7 and 300 K using a cryostat designed solely for XFH experiments (Pretech Co. Ltd., Type XFME-RR4K) at BL39XU of the SPring-8, Sayo, Japan. The sample was placed on a rotatable table in the cryostat head part. The incident x-rays were focused with a size of $0.3\times0.3$ mm$^2$ on the (001) surface of the sample. The measurement was performed in inverse mode \cite{HayashiJPCM} by changing the azimuthal angle $0^\circ\le\phi\le360^\circ$ in steps of $\sim0.35^\circ$ of the cryostat head and the incident angle $0^\circ\le\phi\le75^\circ$ in steps of $1^\circ$ of the whole cryostat. The Yb $L\alpha$ fluorescent x-rays with an energy of 7.414 keV were collected using an avalanche photodiode detector $via$ a cylindrical graphite crystal energy-analyzer. The XFH signals were recorded at $E_0$ of 8.939 and 8.947 keV as indicated by arrows in Fig. \ref{YbXANES}. 

The holographic oscillation data were obtained by subtracting the backgrounds from the fluorescent x-ray intensities and normalized them with the incident x-ray intensities measured with an ion chamber. Extensions of the holographic data were carried out using the measured x-ray standing wave lines in the hologram. Since the holograms are obtained at a single energy in the present study, a usual Fourier transform-like analysis produces unphysical twin and false images owing to too less input experimental data for the requested unknown atomic configurations. Thus, we employed a sophisticated analysis of a "scattering pattern matrix extraction algorithm using an $L_{\rm 1}$ regularized linear regression" (SPEA-L1) by Matsushita \cite{Matsushita}, which is based on inverse problem and represents a sparse modeling approach to the experimental holographic data.

The holographic oscillation $\chi({\bvec k})$ is given as, 
\begin{equation}
\chi({\bvec k})=-\int g({\bvec r})f(\theta_{{\bvec r},{\bvec k}})\cos(kr-{\bvec k}\cdot{\bvec r}){\rm d}{\bvec r},
\label{original}
\end{equation}
where $g({\bvec r})$ is the atomic distribution function and $f(\theta_{{\bvec r},{\bvec k}})$ is the atomic form factor at an angle between {\bvec r} and {\bvec k}, $\theta_{{\bvec r},{\bvec k}}$. when voxels are introduced for describing $g({\bvec r})$, Eq. (\ref{original}) is modified as,
\begin{equation}
\chi({\bvec k}_j)=-\sum g({\bvec r}_i)f(\theta_{{\bvec r}_i,{\bvec k}_j})\cos(k_jr_i-{\bvec k}_j\cdot{\bvec r}_i),
\label{voxel}
\end{equation}
where $i$ and $j$ are the numbers of the voxels and pixel of the holograms, respectively. Since $g({\bvec r}_i)$ is sparse in real space, a $L_{\rm 1}$-regularized regression is applicable. To obtain $g({\bvec r}_i)$, its evaluation function is given by,
\begin{equation}
E=\sum_j|\chi({\bvec k}_j)-\hat{\chi}({\bvec k}_j)|^2+\lambda\sum_i|g({\bvec r}_i)|,
\end{equation}
where $\hat{\chi}({\bvec k}_j)$ is the experimental hologram and $\lambda$ is a penalty parameter. For the present analysis, the voxel size was set to be 0.01 nm cubic in the total range of $\pm0.6$ nm for each direction from the central Yb atoms.

\begin{figure*}
\begin{center}
\includegraphics[width=130mm]{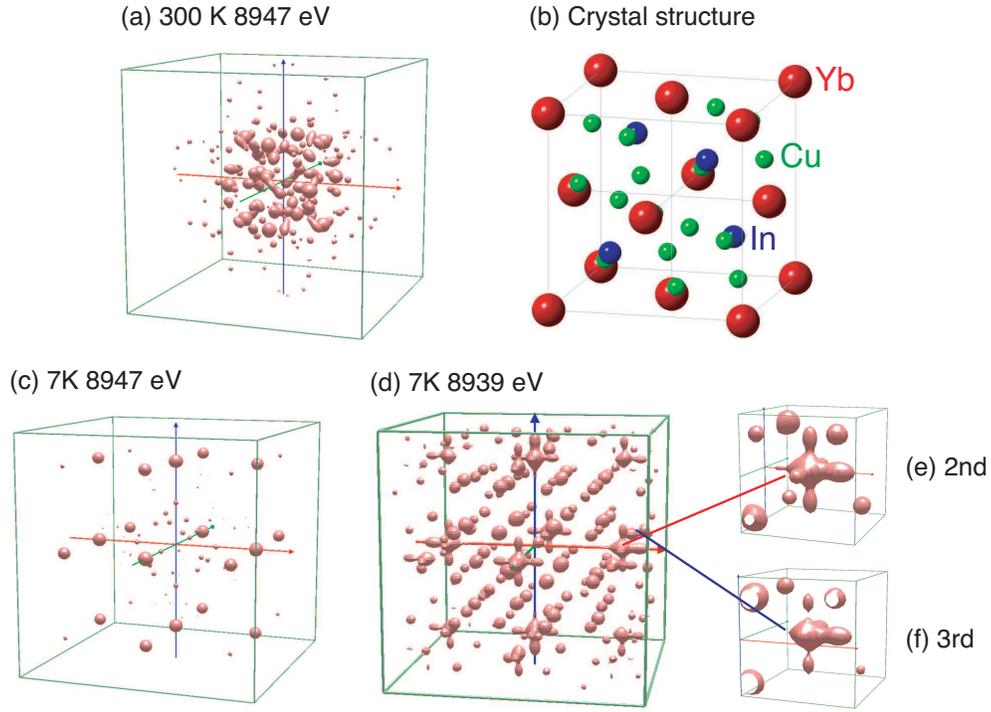}
\caption{\label{3DimageYb}3D atomic images measured at (a) 300 and (c) 7 K with $E_0=8.947$ keV, and (d) those at 7 K at $E_0=8.939$ keV. For the clarity for (d), 2nd and 3rd neighboring Yb atoms are enlarged in (e) and (f), respectively. (b) Crystal structure \cite{KojimaJMMM} for the reference.} 
\end{center}
\end{figure*}

Figure \ref{3DimageYb}(a) shows 3D atomic images measured at 300 K and $E_0=8.947$ keV, where mainly Yb$^{3+}$ ions emit fluorescent x-rays. The threshold value is set at 20\% of the maximum intensity in (c). For the reference, the crystal structure is given in (b), where large, middle, and small balls indicate Yb, In, and Cu atoms, respectively. At a glance, only the first-neighboring Yb atoms are observed. 

Figure \ref{3DimageYb}(c) shows 3D images measured at 7 K with the same $E_0$ value, where both the Yb$^{3+}$ and Yb$^{2+}$ ions are excited. As clearly seen in the figure, only the Yb ions are observed due probably to the large electron numbers of Yb rather than other elements, and hereafter only the Yb ions will be discussed. Interestingly, second- and third-neighboring Yb atoms become clear with decreasing temperature, while the first-neighboring Yb images are hardly seen. Such a feature was frequently seen in XFH results of impurity doping \cite{HosokawaInGaSb,HosokawaZnMnTe}, which originates from positional fluctuations caused by unusual atoms such as impurities or Yb$^{2+}$ atoms.

Figure \ref{3DimageYb}(c) shows 3D images measured at 7 K with $E_0=8.939$ keV, where only the Yb$^{2+}$ ions are excited. To observe the positional distortions in detail, the threshold is lowered at 10\%. The first-neighboring Yb atoms are again invisible due probably to large positional fluctuations. In contrast to (c), the second- and third-neighboring Yb atoms are highly deformed, and their enlarged figures are given in (e) and (f), respectively. As clearly seen in (d)--(f), both the second- and third-neighboring atomic images have tails in the $x$, $y$, and $z$ directions, and they are they are stronger in the distant directions with respect to the central Yb$^{2+}$ atom. 

\begin{figure*}
\begin{center}
\includegraphics[width=120mm]{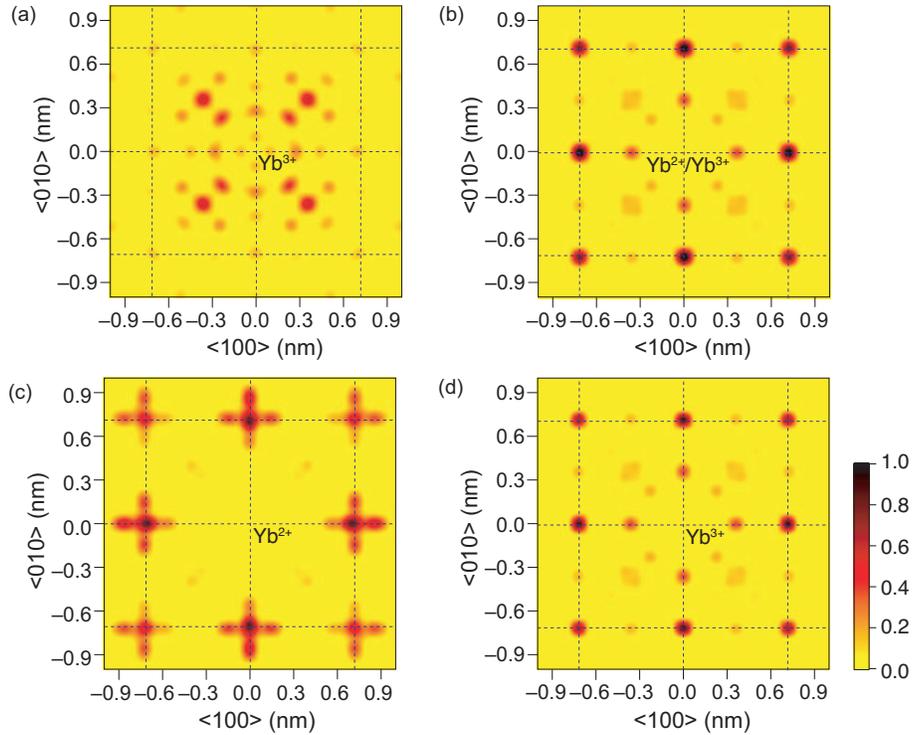}
\caption{\label{2DimageYb}2D atomic images on the (001) plane at $z=0$ measured at (a) 300 and (b) 7 K with $E_0=8.947$ keV. (c) Those at 7 K at $E_0=8.939$ keV. (d) Estimated pure Yb$^{3+}$ contributions obtained from the contrast between (b) and (c).}
\end{center}
\end{figure*}

To observe the atomic images further and quantitatively, Fig. \ref{2DimageYb} shows those on the (001) plane at $z=0$, where mostly the Yb$^{3+}$ ions are located in the C15b structure. The image intensities are indicated by the color bar besides the figures. Figure \ref{2DimageYb}(a) corresponds to Fig. \ref{3DimageYb}(a) measured at 300 K with $E_0=8.947$ keV. As mentioned above, the stronger images of about 0.4 are located at the first-neighboring positions although unphysical artifacts are seen around the first-neighboring atoms. In addition, weak atomic images of about 0.1 are detected at the second- and third-neighboring positions.

Figure \ref{2DimageYb}(b) corresponds to Fig. \ref{3DimageYb}(c) measured at 7 K with $E_0=8.947$ keV, indicating atomic images around the Yb$^{3+}$ and Yb$^{2+}$ ions. Mostly $fcc$ sublattice is seen in the images besides weak images of about 0.2 at the first-neighboring positions. Figure \ref{2DimageYb}(c) corresponds to Fig. \ref{3DimageYb}(d) measured at 7 K with $E_0=8.939$ keV, indicating atomic images around only the Yb$^{2+}$ ions. The  images of second- and third-neighboring Yb ions are distorted along the cross directions of $\langle100\rangle$ and $\langle010\rangle$. 

Since the ratio of the Yb$^{2+}$ was determined by various methods to be 0.15--0.26 \cite{Zhuang,Sato,Yamaoka}, atomic arrangements around the pure Yb$^{3+}$ ions can be estimated by assuming that 20\% of (b) is contributed by (c), and the obtained pure Yb$^{3+}$ atomic images are shown in Fig. \ref{2DimageYb}(d). Since the fraction of the Yb$^{2+}$ is not large,  (d) is very similar to (b), and the error in the Yb$^{3+}$ fraction affect the estimated result very slightly. Thus, it is concluded that valence-selective XFH experiment provides a clear difference in local atomic arrangements, i.e., the reconstructed images around Yb$^{3+}$ show a clear $fcc$ structure as observed by diffraction experiments, whereas those around Yb$^{2+}$ have curious cross features. 

\begin{figure}
\begin{center}
\includegraphics[width=40mm]{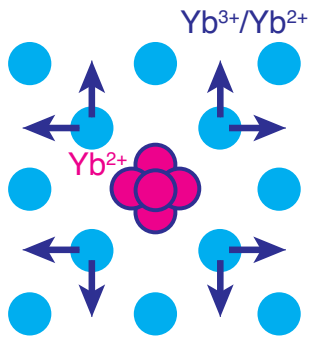}
\caption{\label{YbInCu4model} A reasonable model around the Yb$^{2+}$ ions obtained from the present XFH results.}
\end{center}
\end{figure}

Based on the experimental results, a structural model is proposed around the Yb$^{2+}$ ions as shown in Fig. \ref{YbInCu4model}. Since the ionic radius of Yb$^{2+}$ is larger than that of Yb$^{3+}$ by about 17\% \cite{Shannon} while the increase in the lattice constant is only about 0.15\% \cite{KojimaJPSJ}, the Yb$^{2+}$ ions may hardly stay at the lattice positions. The shifts cause by avoiding the first-neighboring Tb atoms, i.e., $\langle100\rangle$, $\langle010\rangle$, or $\langle001\rangle$ direction, as shown in the shifted balls of the figure, which causes curious hexagonal-cross atomic images for the second- and third-neighboring Yb atoms. By moving the central Yb$^{2+}$ ion, this atom pushes the first-neighboring atoms towards the same directions to avoid the second- and third-neighboring atoms as indicated by the arrows in the figure. Therefore, the positional fluctuations of the first-neighboring atoms are very large.

In summary, a valence-selective XFH experiment was applied to a YbInCu$_4$ valence transition material to investigate the local atomic arrangements around the Yb$^{2+}$ and Yb$^{3+}$ ions in this material. A large difference was observed, indicating an excellent feasibility for obtaining the valence-selective structural information, which is not easy by usual diffraction and XAFS measurements.

%\section{Results}

%\section{Discussion}

%\section{Summary}

%\begin{acknowledgments}
The authors thank Professor Ichiro Akai and Professor Masato Okada for useful information on the sparse modeling. The XFH experiments were performed at the beamline BL39XU in the SPring-8 with the approval of the Japan Synchrotron Radiation Research Institute (JASRI) (Proposal No. 2015A1005 and 2018A1214). This works was supported by JSPS Grant-in-Aid for Scientific Research (B) (No. 17H02814), those on Innovative Areas "3D Active-Site Science" (Nos. 26105006 and 26105013) and "Sparse Modeling" (No. 16H01553), and by JST CREST (No. JPMJCR1861). JRS gratefully acknowledges a financial support as Overseas Researcher under a JSPS fellowship (No. P16796).
%\end{acknowledgments}


\begin{references}

\bibitem{Felner86}
I. Felner and I. Nowik, 
Phys. Rev. B {\bf 33}, 617 (1986).

\bibitem{Felner87}
I. Felner et al., %I. Nowik, D. Vaknin, U. Potzel, J. Moser, G. M. Kalvius, G. Wortmann, G. Schmiester, G. Hilscher, E. Gratz, C. Schmitzer, N. Pillmayr, K. G. Prasad, H. de Waard, and H. Pinto,
Phys. Rev. B {\bf 35}, 6956 (1987).

\bibitem{FelnerJMMM}
I. Felner and I. Nowik, 
J. Mag. Mag. Mater. {\bf 63\&64}, 615 (1987).

\bibitem{Nowik}
I. Nowik et al., %I. Felner, J. Voiron, J. Beille, A. Najib, E. du Tremolet de Lacheisserie, and G. Gratz, 
Phys. Rev. B {\bf 37}, 5633 (1988).

\bibitem{KojimaJMMM}
K. Kojima et al., %H. Hayashi, A. Minami, Y. Kasamatsu, and T. Hihara, 
J. Mag. Mag. Mater. {\bf 81}, 267 (1989).

\bibitem{KojimaJPSJ}
K. Kojima et al., %Y. Nakai, T. Suzuki, H. Asano, F. Izumi, T. Fujita, and T. Hihara, 
J. Phys. Soc. Jpn. {\bf 59}, 792 (1990).

\bibitem{Kindler}
B. Kindler et al., %D. Finsterbusch, R. Graf, F. Ritter, W. Assmus, and B L\"{u}chi, 
Phys. Rev. B {\bf 50}, 704 (1994).

\bibitem{Lawrence}
J. M. Lawrence et al., %G. H. Kwei, J. L. Sarrao, Z. Fisk, D. Mandrus, and J. D. Thompson, 
Phys. Rev. B {\bf 54}, 6011 (1996).

\bibitem{Moriyoshi} 
C. Moriyoshi et al., %S. Shimomura, K. Itoh, K. Kojima, and K. Hiraoka, 
J. Mag. Mag. Mater. {\bf 260}, 206 (2003).

\bibitem{Zhuang}
T. Zhuang et al., %K. Hiraoka, M. Kurisu, K. Konishi, T. Kamimori, and I. Nakai, 
JPS Conf. Proc. {\bf 3}, 011069 (2014).

\bibitem{Sato}
H. Sato et al., %K. Shimada, M. Arita, K. Hiraoka, K. Kojima, Y. Takeda, K. Yoshikawa, M. Sawada, M. Nakatake, H. Namatame, M. Taniguchi, Y. Takata, E. Ikenaga, S. Shin, K. Kobayashi, K. Tamasaku, Y. Nishino, D. Miwa, M. Yabashi, and T. Ishikawa, 
Phys. Rev. Lett. {\bf 93}, 246404 (2004).

\bibitem{Yamaoka}
H. Yamaoka et al., %N. Tsujii, K. Yamamoto, A. M. Vlaicu, H. Oohashi, H. Yoshikawa, T. Tochio, Y. Ito, A Chainani, and S. Shin, 
Phys. Rev. B {\bf 78}, 045127 (2008).

\bibitem{UtsumiPRB}
Y. Utsumi et al., %H. Sato, H. Kurihara, H. Maso, K. Hiraoka, K. Kojima, K. Tobimatsu, T. Ohkochi, S. Fujimori, Y. Takeda, Y. Saitoh, K. Mimura, S. Ueda, Y. Yamashita, H. Yoshikawa, K. Kobayashi, T. Oguchi, K. Shimada, H. Namatame, and M. Taniguchi, 
Phys. Rev. B {\bf 84}, 115143 (2011).

\bibitem{Shannon}
R. D. Shannon, 
Acta Cryst. A{\bf 32}, 751 (1976).

\bibitem{UtaumiJJAP}
Y. Utsumi et al., %H. Sato, C. Moriyoshi, Y. Kuroiwa, H. Namatame, M. Taniguchi, K. Hiraoka, K. Kojima, and K. Sugimoto, 
Jpn. J. Appl. Phys. {\bf 50}, 05FC10 (2011). 

\bibitem{Tsutsui}
S. Tsutsui et al., %K. Sugimoto, R. Tsunoda, Y. Hirose, T. Mito, R. Settai, and M. Mizumaki, 
J. Phys. Soc. Jpn. {\bf 85}, 063602 (2016). 

\bibitem{Stellhorn}
J. R. Stellhorn et al., %S. Hosokawa, N. Happo, H. Tajiri, T. Matsushita, K. Kaminaga, T. Fukumura, T. Hasegawa, and K. Hayashi,
J. Appl. Crystallogr. {\bf 50}, 1583 (2017).

\bibitem{Ang}
A. K. R. Ang et al., %T. Matsushita, Y. Hashimoto,  N. Happo, Y. Yamamoto, M. Mizuguchi, A. Sato-Tomita, N. Shibayama, Y. C. Sasaki, K. Kimura, M. Taguchi, H. Daimon, and K. Hayashi, 
Phys. Status Solidi B 255, 1800100 (2018).

\bibitem{Tegze}
M. Tegze and G. Faigel, Europhys. Lett. {\bf 16}, 41 (1991).

\bibitem{HayashiJPCM}
K. Hayashi et al., %N. Happo, S Hospkawa, W. Hu, and T. Matsushita, 
J. Phys.: Condens. Matter {\bf 24}, 093201 (2012).

\bibitem{Matsushita}
T. Matsushita, 
e-J. Surf. Sci. Nanotech. {\bf 14}, 158 (2016).

\bibitem{HosokawaInGaSb}
S. Hosokawa et al., %N. Happo, T. Ozaki, H. Ikemoto, T. Shishido, and K. Hayashi,
Phys. Rev. B {\bf 87}, 094104 (2013).

\bibitem{HosokawaZnMnTe}
S. Hosokawa, N. Happo, and K. Hayashi, 
Phys. Rev. B {\bf 80}, 134123 (2009).

\end{references}
\end{document}